\documentclass{tlp}    

\usepackage{times}
\usepackage[german,english]{babel}
\usepackage{algorithm}
\usepackage{algorithmic}
\usepackage{epsfig}
\usepackage{xspace}
\usepackage{url}
\usepackage{harvard}

\newcommand{\nop}[1]{}

\newcommand{\dlv}{{\bf\small{DLV}}\xspace}
\newtheorem{example}{Example}

\newtheorem{theorem}{Theorem}

\newtheorem{definition}{Definition}
\newtheorem{proposition}{Proposition}

\newtheorem{lemma}{Lemma}

\newenvironment{theorem*}[2]%
{\begin{trivlist} \item[] {\bf
#1~\protect{\ref{#2}}}\it}{\end{trivlist}}

\def\Or{\vee}

\def\U{{U_\gp}}

\def\DOL{{\cal DOL}}

\def\O{{\cal O}}
\def\gp{{\cal P}}

\def\+{\hspace*{0.5cm}}
\def\-{\+\+}

\def\DOL{{\cal DOL}}

\def\BP {B_{\gp}}

\def\derives{\leftarrow}

\newcommand{\tuple}[1]{\langle#1\rangle}

\newcommand{\SigmaP}[1]{{\Sigma}_{#1}^{P}}
\newcommand{\PiP}[1]{{\Pi}_{#1}^{P}}

\newcommand{\NP}{\mbox{${\rm N\!P}$}}

\newcommand{\CONP}{\mbox{${\rm coN\!P}$}}

\newcommand{\irule}[1]{\ensuremath{\mathtt{#1}}}

\newcommand{\tneg}{\ensuremath{\neg}}
\newcommand{\dneg}{\ensuremath{\mathtt{not}}}

\def\punto{\hspace*{\fill}{\rule[-0.5mm]{1.5mm}{3mm}}}
\def\DLPI{{\rm DLP\mbox{$^{<}$}}}

\def\DOL{{\cal DOL}}

\def\bldl{\smallskip\[\bf\begin{array}{ll}}
\def\cldl{\vspace{-0.4cm}\[\bf\begin{array}{ll}}
\def\eldl{\end{array}\]\rm}

\begin{document}
\bibliographystyle{dcu}

\title{Disjunctive Logic Programs with Inheritance}
\author[F. Buccafurri, W. Faber and N. Leone]
{FRANCESCO BUCCAFURRI\\
DIMET -- Universit{\`a} di Reggio Calabria
89100, Reggio Calabria, Italia\\
\email{bucca@ns.ing.unirc.it}
\and
WOLFGANG FABER\\
Institut f{\"u}r Informationssysteme,
Technische Universit{\"a}t Wien,
1040 Vienna, Austria\\
\email{faber@kr.tuwien.ac.at}
\and
NICOLA LEONE\\
Dipartimento di Matematica,
Universit{\`a} degli Studi della Calabria,
87030 Rende (CS), Italia\\
\email{leone@unical.it}}

\maketitle

\begin{abstract}
The paper proposes a new knowledge representation language, called $\DLPI$,
which extends disjunctive logic programming (with strong negation)
by inheritance.
The addition of inheritance enhances the knowledge modeling
features of the language providing a natural representation
of default reasoning with exceptions.\\
A declarative model-theoretic semantics of $\DLPI$ is provided,
which is shown to generalize the Answer Set Semantics
of disjunctive logic programs.\\
The knowledge modeling features of the language are illustrated
by encoding classical nonmonotonic problems in $\DLPI$.\\
The complexity of $\DLPI$ is analyzed, proving that inheritance does
not cause any computational overhead, as reasoning in $\DLPI$ has
exactly the same complexity as reasoning in disjunctive logic
programming.  This is confirmed by the existence of an efficient
translation from $\DLPI$ to plain disjunctive logic programming.
Using this translation, an advanced KR system supporting the $\DLPI$
language has been implemented on top of the \dlv system and has
subsequently been integrated into \dlv.
\end{abstract}

\section{Introduction}
Disjunctive logic programs
are logic programs where disjunction is allowed in the heads of the rules
and negation as failure (NAF) may occur in the bodies of the rules.
Such programs are now widely recognized as a valuable tool for
knowledge representation and commonsense reasoning
\cite{bara-gelf-94,lobo-etal-92,gelf-lifs-91}.
One of the attractions of disjunctive logic programming
is its ability to naturally model incomplete
knowledge \cite{bara-gelf-94,lobo-etal-92}.
The need to differentiate between atoms which are false because
of the failure to prove them true (NAF, or CWA negation)
and atoms the falsity of which is explicitly provable
led to extend disjunctive logic programs by strong negation
\cite{gelf-lifs-91}.
Strong negation, permitted also in the heads of rules,
further enhances the knowledge modeling features of the language,
and its usefulness is widely acknowledged in the literature
\cite{alfe-pere-92,bara-gelf-94,kowa-sadr-90,alfe-etal-96,saka-inou-96,alfe-etal-98a}.
However, it does not allow to represent default reasoning with
exceptions in a direct and natural way.
Indeed, to render  a default rule $r$ {\em defeasible},
$r$ must at least be equipped with an extra negative literal,
which ``blocks'' inferences from $r$ for abnormal instances \cite{gelf-son-97}.
For instance, to encode the famous nonmonotonic reasoning (NMR) example stating
that birds {\em normally} fly while penguins do not fly, one should 
write%
\footnote{\dneg{} and \tneg{} denote the weak negation symbol
and the strong negation symbol, respectively.}
the rule
\[\irule{fly(X) \derives bird(X),\ \dneg{}\ \tneg fly(X).}\]
along with the fact
\[\irule{\tneg fly(penguin).}\]

This paper proposes an extension of disjunctive logic programming
by inheritance, called $\DLPI$. The addition of inheritance
enhances the knowledge modeling features of the language. Possible
conflicts are solved in favor of the rules which are ``more
specific'' according to the inheritance hierarchy. This way, a
direct and natural representation of default reasoning with
exceptions is achieved (e.g., defeasible rules do not need to be
equipped with extra literals as above -- see section \ref{KR}).

The main contributions of the paper are the following:
\begin{itemize}
\item
We formally define the $\DLPI$ language, providing a declarative
model theoretic semantics of $\DLPI$, which
is shown to generalize the Answer Set Semantics
of \cite{gelf-lifs-91}.
\item
We illustrate the knowledge modeling features of the language
by encoding classical nonmonotonic problems in $\DLPI$.
Interestingly, $\DLPI$ also supplies a very natural representation of
frame axioms.
\item
We analyze the computational complexity of reasoning over $\DLPI$
programs.
Importantly,
while inheritance enhances the knowledge modeling ability of disjunctive logic
programming, it does not cause any computational overhead,
as reasoning in $\DLPI$ has exactly the same complexity
as reasoning in disjunctive logic programming.
\item
We compare $\DLPI$ to related work proposed in the literature. In
particular, we stress the differences between $\DLPI$ and
Disjunctive Ordered Logic ($\DOL$)
\cite{bucc-etal-98a,bucc-etal-99c}; we point out the relation to
the Answer Set Semantics of \cite{gelf-lifs-91}; we compare $\DLPI$
with prioritized disjunctive logic programs \cite{saka-inou-96};
we analyze its relationships to inheritance networks \cite{tour-86}
and we discuss the possible application of $\DLPI$
to give a formal semantics to updates of logic programs.
\cite{alfe-etal-98b,mare-trus-94,leon-etal-95b}.
\item
We implement a $\DLPI$ system.
To this end, we first design an efficient translation
from $\DLPI$ to plain disjunctive logic programming.
Then, using this translation, we implement a
$\DLPI$ evaluator on top of the \dlv system
\cite{eite-etal-98a}.
It is part of \dlv and can be freely retrieved from \cite{dlvi-web}.
\end{itemize}

The sequel of the paper is organized as follows.
The next two sections provide a formal definition of $\DLPI$;
in particular, its syntax is given in Section
\ref{syntax} and its semantics is defined in Section \ref{sec:semantics}.
Section \ref{KR} shows the use of $\DLPI$ for knowledge representation
and reasoning, providing a number of sample $\DLPI$ encodings.
Section \ref{sec:complexity} analyzes the computational complexity
of the main reasoning tasks arising in the framework of $\DLPI$.
Section \ref{sec:relatedwork} discusses related work.
The main issues underlying the implementation
of our $\DLPI$ system are tackled in Section \ref{sec:implementation},
and our conclusions are drawn in Section \ref{sec:conclusion}.

\section{Syntax of $\DLPI$}\label{syntax}
This section provides a formal description of
syntactic constructs of the language.

Let the following disjoint sets be given: a set $\cal V$
of {\em variables}, a set
$\Pi$ of {\em predicates}, a set
$\Lambda$ of {\em constants}, and
a finite partially ordered set of symbols
$(\cal O, <)$,
where $\cal O$ is a set of strings, called {\em object identifiers},
and $<$ is a strict partial order
(i.e., the relation $<$ is:
(1) irreflexive -- $c \not < c\;\; \forall c\in \cal O$, and
(2) transitive -- $a<b \wedge b<c \Rightarrow a<c\;\; \forall a,b,c\in \cal O$).

A $term$ is either a constant in $\Lambda$ or a variable
in $\cal V$.%
\footnote{Note that function symbols
are not considered in this paper.}

An $atom$ is a construct of the form
$a(t_{1},..., t_{n})$, where $a$ is a $predicate$ of arity $n$
in $\Pi$ and $t_{1},..., t_{n}$ are terms.

A $literal$ is either a $positive~literal$ $p$ or a
$negative~literal$ $\tneg ~p$, where $p$ is an atom ($\tneg$ is the $strong~negation$ symbol).
Two literals are $complementary$ if they are of the form $p$ and $\tneg
p$, for some atom $p$.

Given a literal $L$, $\tneg .L$ denotes its complementary literal.
Accordingly, given a set $A$ of literals,
$\tneg .A$ denotes the set \{$\tneg .L~|~L \in A$\}.

A {\em  rule} $r$ is an expression of the form

$\quad
a_1\Or\cdots\Or
        a_n \derives b_1,\cdots, b_k, \dneg\;
        b_{k+1},\cdots, \dneg\; b_m \ \bigotimes \quad\quad n\geq 1,\ m\geq 0
$
\\
where $a_1,\cdots ,a_n,b_1,\cdots ,b_m$ are literals,
$\dneg$ is the {\em negation as failure} symbol
and $\bigotimes$ is either (1) the symbol '.' or (2)
the symbol '!'.
In case (1) $r$ is a {\em defeasible rule},
in case (2) it is a {\em strict rule}.

The disjunction $a_1\Or\cdots\Or a_n$ is the {\em head} of $r$, while
the conjunction $b_1,$ $\ldots,$ $b_k,$ $\dneg\; b_{k+1},$ $\ldots,$ $\dneg\; b_m$ is the {\em
body} of $r$.
$b_1, ... , b_k$ is called the
{\em positive part} of the body of $r$ and
$\dneg\;b_{k+1},... ,\dneg\; b_ m$ is called the
{\em NAF (negation as failure) part} of the body of $r$.
We often denote the sets of literals appearing in the head, in the
positive, and in the NAF part of the body of a rule
$r$ by $Head(r)$, $Body^+(r)$, and $Body^-(r)$, respectively.

If the body of a rule $r$ is empty, then $r$ is called {\em fact}. The
symbol '$\derives$' is usually omitted from facts.

An {\em object} $o$ is a pair $\tuple{oid(o), \Sigma(o)}$, where $oid(o)$ is an
object identifier in $\cal O$ and $\Sigma(o)$ is a (possibly empty) set of
rules.

A {\em knowledge base} on $\cal O$ is a set of objects, one for each element
of $\cal O$.

Given a knowledge base $\cal K$ and an object identifier $o \in \cal O$, the
{\em $\DLPI$ program for} $o$ ({\em on} $\cal K$)
is the set of objects
$\gp = \{ (o' , \Sigma(o' )) \in {\cal K} \ | \  o=o' \ or \ o <  o'  \}$.

The relation $<$ induces a partial order on $\gp$ in the obvious way,
that is, given $o_i = (oid(o_i),\Sigma(o_i))$ and $o_j= (oid(o_j),
\Sigma(o_j))$,
$o_i < o_j$ iff $oid(o_i) < oid(o_j)$ (read "$o_i$ is more specific
than $o_j$").

A term, an atom, a literal, a rule, 
or program is {\em ground} if no variable appears in it.

Informally, a knowledge base can be viewed as a set of
{\em objects} embedding the definition of their
properties specified through disjunctive logic rules,
organized in an IS-A (inheritance) hierarchy
(induced by the relation $<$).
A program $\gp$ for an object $o$ on a knowledge base $\cal K$
consists of the portion of $\cal K$ "seen" from $o$ looking up in
the IS-A hierarchy. Thanks to the inheritance mechanism, $\gp$
incorporates the knowledge explicitly defined for $o$ plus the
knowledge inherited from the higher objects.

If a knowledge base admits a {\em bottom} element (i.e., an object
less than all the other objects, by the relation $<$), we usually
refer to the knowledge base as ``program'', since it is equal to the
program for the bottom element.

Moreover, we represent
the {\em transitive reduction} of
the relation $<$ on the objects.%
\footnote{$(a,b)$ is in the
transitive reduction of $<$ iff $a < b$ and
there is no $c$ such that $a < c$ and $c < b$.}
An object $o$ is denoted as $oid(o) : o_1, \ldots, o_n \; \Sigma(o)$%
\footnote{The set $\Sigma(o)$ is denoted without commas as separators.}
, where
$(oid(o),o_1),$ $\ldots$, $(oid(o),o_n)$ are exactly those pairs of the transitive reduction of $<$,
in which the first object identifier is $oid(o)$.  $o$ is referred to as {\em sub-object} of
$o_1, \ldots, o_n$.

\begin{example}\label{ex-1}
{
Consider the following program $\gp$:
\begin{tabbing}
\quad\quad \irule{o_2:o_1}  \= \{ \= p \kill
\quad\quad \irule{o_1}      \> \{ \> \irule{a \Or \tneg b \leftarrow c, \dneg \ d.} \quad \irule{e \derives b!} \ \} \\
\quad\quad \irule{o_2:o_1}  \>  \{ \> \irule{b.} \quad \irule{\tneg a \Or c.} \quad \irule{c \leftarrow b.} \}
\end{tabbing}
$\gp$ consists of two objects \irule{o_1} and \irule{o_2}.
\irule{o_2} is a sub-object of \irule{o_1}.
According to the convention illustrated above, the knowledge base on which
$\gp$ is defined coincides with $\gp$, and the object
for which $\gp$ is defined is \irule{o_2} (the bottom object).
}
$\punto$
\end{example}

\section{Semantics of $\DLPI$}\label{sec:semantics}
In this section we assume that a knowledge base $\cal K$ is given and
an object $o$ has been fixed.
Let $\gp$ be the $\DLPI$ program for $o$ on $\cal K$.
The {\em Universe} $\U$ of $\gp$ is the set of all constants
appearing in the rules. 
The {\em Base} $\BP$ of $\gp$ is the set of all
possible ground literals constructible from
the predicates appearing in the rules of $\gp$ and
the constants occurring in $\U$.
Note that, unlike in traditional logic programming the Base of
a $\DLPI$ program contains both positive and negative literals.
Given a rule 
$r$  occurring in $\gp$,
a {\em ground instance} of $r$ is a rule obtained from $r$
by replacing every variable $X$ in $r$ by $\sigma(r)$, where
$\sigma$ is a mapping from the variables occurring in $r$ to
the constants in $\U$.
We denote by $ground( \gp )$ the (finite) multiset of
all instances of
the rules 
occurring in $\gp$.
The reason why $ground(\gp)$
is a multiset is that a rule may appear in several
different objects of $\gp$, and we require the respective ground
instances be distinct.
Hence, we can define a function $obj\_of$ from
ground instances of rules in $ground(\gp)$
onto the set $\cal O$ of the object identifiers, associating
with a ground instance ${\overline r}$ of $r$
the (unique) object of $r$.

A subset of ground literals in $\BP$ is said to be {\em consistent} if
it does not contain a pair of complementary literals.
An {\em interpretation} $I$ is a consistent subset of $\BP$.
Given an interpretation $I \subseteq \BP$,
a ground literal (either positive
or negative) $L$ is {\em true}
w.r.t. $I$ if $L \in I$ holds.
$L$ is {\em false} w.r.t. $I$ otherwise.

Given a rule $r \in ground( \gp )$, the head of
$r$ is {\em true} in $I$ if
at least one literal of the head is true w.r.t $I$.
The body of $r$ is {\em true} in $I$ if:
(1) every literal in $Body^+(r)$ is true w.r.t. $I$, and
(2) every literal in $Body^-(r)$ is false w.r.t. $I$.
A rule $r$ is {\em satisfied} in $I$ if
either the head of $r$ is true in $I$ or the
body of $r$ is not true in $I$.

Next we introduce the concept of a {\em model}
for a $\DLPI$-program.
Different from traditional logic programming, the
notion of satisfiability of rules
is not sufficient for this goal,
as it does not take into account the presence of explicit contradictions.
Hence, we first present some preliminary definitions.

Given two ground rules $r_1$ and $r_2$
we say that $r_1$ {\em threatens} $r_2$ {\em on}
a literal $L$ if (1)
$\tneg . L \in Head(r_1)$ and $L \in Head(r_2)$,
(2) $obj\_of(r_1) < obj\_of(r_2)$
and (3) $r_2$ is defeasible,

\begin{definition}\label{ov}
{\em
Given an interpretation $I$ and two ground rules
$r_1$ and $r_2$ such that $r_1$ threatens $r_2$ on $L$ we say that
$r_1$ {\em overrides} $r_2$ {\em on} $L$ {\em in} $I$ if:
(1) $\tneg. L \in I$, and
(2) the body of $r_2$ is true in $I$.

A (defeasible) rule $r \in ground(\gp)$ is {\em overridden in}
$I$ if for each $L \in Head(r)$ there
exists $r_1 \in ground(\gp)$ such that
$r_1$ overrides $r$ on $L$ in $I$.
}
$\punto$
\end{definition}

Intuitively, the notion of overriding allows us to solve conflicts
arising between rules with complementary heads.
For instance, suppose that both $a$ and $\tneg a$ are derivable in $I$
from rules $r$ and $r'$, respectively.
If $r$ is more specific than $r'$ in the inheritance hierarchy
and $r'$ is not strict,
then $r'$ is overridden, meaning that $a$ should be preferred to $\tneg a$
because it is derivable from a more trustable rule.

Observe that, by definition of overriding, strict rules cannot
be overridden, since they are never threatened.

\begin{example}\label{ex-2}
{
Consider the program $\gp$ of Example \ref{ex-1}.
Let \irule{I = \{ \tneg a, b, c, e \}} be an interpretation.
Rule \irule{\tneg a \Or c \leftarrow. } in the object \irule{o_2} overrides
rule \irule{a \Or \tneg b \leftarrow c, \dneg \ d.} in \irule{o_1}
on the literal \irule{a} in \irule{I}.
Moreover, rule \irule{b  \leftarrow.} in \irule{o_2} overrides
rule \irule{a \Or \tneg b \leftarrow c, \dneg \ d.} in \irule{o_1}
on the literal \irule{\tneg b} in \irule{I}.
Thus, the rule
\irule{a \Or \tneg b \leftarrow c, \dneg \ d.} in \irule{o_1}
is overridden in \irule{I}.
}
$\punto$
\end{example}

\begin{example}\label{ex-5}
{
Consider the following program $\gp$:
\begin{tabbing}
\quad\quad \irule{o_2:o_1} \= \{ \= p \kill
\quad\quad \irule{o_1}     \> \{ \> \irule{\tneg a!} \quad \irule{\tneg b.} \ \} \\
\quad\quad \irule{o_2:o_1} \> \{ \> \irule{a \derives \dneg \ b.} \quad \irule{b \derives \dneg \ a.}  \}
\end{tabbing}
Consider now the interpretations \irule{M_1 = \{a, \tneg b \}} and 
\irule{M_2 = \{b, \tneg a \}}. While the rule \irule{\tneg b.} is overridden
in \irule{M_2}, the rule \irule{\tneg a!} cannot be overridden since
it is a strict rule. Due to overriding, strict rules and defeasible
rules are quite different from the semantic point of view. In our
example, the overriding mechanism allows us to invalidate the
defeasible rule \irule{\tneg b.} in favor of the more trustable
one \irule{b \derives \dneg \ a.} (w.r.t.\ the interpretation \irule{M_2}). In
words, the defeasible rule is invalidated in \irule{M_2} because of a
more specific contradictory rule and no inconsistency is
generated. In other words, it is possible to find an
interpretation containing the literal \irule{b} (i.e., stating an
exception for the rule \irule{\tneg b.}) such that all the rules
of $\gp$ are either satisfied or overridden (i.e., invalidated) in
it. Such an interpretation is just \irule{M_2}. This cannot happen for
the strict rule \irule{\tneg a!}. Indeed, no interpretation
containing the literal \irule{a} (i.e., stating an exception for the
strict rule) can be found which satisfies all non-overridden rules of $\gp$. }
$\punto$
\end{example}
In the example above we have implicitly used the notion of {\em
model} for a program $\gp$ that we next formally provide. A model
for a program is an interpretation satisfying all its non
overridden rules.

\begin{definition}\label{model}
{\em
Let $I$ be an interpretation for $\gp$.
$I$ is a {\em model for} $\gp$ if every rule
in $ground(\gp)$ is satisfied or overridden in  $I$.
$I$ is a {\em minimal model for} $\gp$ if no (proper)
subset of $I$ is a model for $\gp$.
}
$\punto$
\end{definition}

Note that strict rules must be satisfied in every model, since they
cannot be overridden.

\begin{example}\label{ex-5bis}
{
It is easy to see that \irule{M_1=\{ a, \tneg b \}} is not a
model for the program $\gp$ of Example \ref{ex-5} since the rule
\irule{\tneg a!} is neither overridden nor satisfied. On the contrary,
\irule{M_2=\{ b, \tneg a \}} is a model for $\gp$, since
\irule{\tneg b.} is overridden by the rule
\irule{b \derives \dneg \ a.} The latter rule is satisfied
since \irule{b \in M_2}. The rule \irule{\tneg a!}
is satisfied as \irule{\tneg a \in M_2} and
the rule \irule{a \derives \dneg \ b.} is satisfied since
both body and head are false w.r.t. \irule{M_2}.
}
\end{example}

Next we define the transformation $G_I$ based on which
our semantics is defined. This transformation applied to
a program $\gp$ w.r.t. an interpretation $I$
output a set of rules $G_I(\gp)$ with no
negation by failure in the body.
Intuitively, such rules are those remaining from $ground(\gp)$
by (1) eliminating the rules overridden in the interpretation $I$, (2)
deleting rules whose NAF part is not "true" in $I$
(i.e., some literal negated by negation as failure occurs in $I$) and
(3) deleting the NAF part of all the remainder rules.
Since the transformation encodes the overriding mechanism,
the distinction between strict rules and defeasible rules in $G_I(\gp)$
is meaningless
(indeed, there is no difference between strict and defeasible rules
except for the overriding mechanism where
the upper rule is required to be defeasible).
For this reason the syntax of rules in $G_I(\gp)$
can be simplified by dropping the symbol $.$ from defeasible rules
and the symbol $!$ from strict rules.
\begin{definition}\label{reduction}
{\em
Given an interpretation $I$ for $\gp$,
the {\em reduction of} $\gp$
{\em w.r.t.} $I$, denoted by
$G_I(\gp)$, is the set of rules
obtained from $ground(\gp)$ by
(1) removing every rule overridden in $I$,
(2) removing every rule $r$ such that
$Body^-(r) \cap I \neq \emptyset$,
(3) removing the NAF part from the bodies of the remaining rules.
}
$\punto$
\end{definition}

\begin{example}\label{ex-3}
{
Consider the program $\gp$ of Example \ref{ex-1}.
Let \irule{I} be the interpretation
\irule{\{\tneg a , b, c, e\}}. As
shown in Example \ref{ex-2}, rule
\irule{a \Or \tneg b \derives c, \dneg \ d.} is overridden in \irule{I}.
Thus,
$G_I(\gp)$ is the set of rules
\{\irule{\tneg a \Or c.} \quad \irule{e \derives b.} \quad \irule{b.} \quad \irule{c \derives b.} \}.
Consider now the interpretation
\irule{M = \{a,b,c,e \}}.
It is easy to see that
$G_{M}(\gp) = \{ \irule{a \Or \tneg b \derives c.} \quad \irule{\tneg a \Or c.}
\quad \irule{e \derives b.} \quad \irule{b.} \quad \irule{c \derives b.} \}$.
}
$\punto$
\end{example}

We observe that the reduction of a program is simply
a set of ground rules. Given a set $S$ of
ground rules, we denote
by $pos(S)$ the
positive disjunctive program
(called the {\em positive version of} $S$), obtained
from $S$ by considering each negative
literal $\tneg p (\bar X)$ as a positive
one with predicate symbol $\tneg p$.

\begin{definition}\label{answerset}
{\em
Let $M$ be a model for $\gp$.
We say that $M$ is a
{\em ($\DLPI$-)answer set} for $\gp$
if $M$ is a minimal model of the positive
version $pos(G_M(\gp))$ of $G_M(\gp)$.
}
$\punto$
\end{definition}

Note that interpretations must be consistent by definition, so
considering $pos(G_M(\gp))$ instead of $G_M(\gp)$ does not lose
information in this respect.

Note that the notion of minimal model of Definition~\ref{model} cannot
be used in Definition~\ref{answerset}, as $G_M$ is a set of rules and
not a $\DLPI$ program.

\begin{example}\label{ex-4}
{
Consider the program $\gp$ of Example \ref{ex-1}:

It is easy to see that the interpretation
\irule{I} of Example \ref{ex-3} is not an answer set
for $\gp$. Indeed, although
\irule{I} is a model for $pos(G_I(\gp))$ it is
not minimal, since the interpretation
\irule{\{b,c,e \}} is a model for $pos(G_I(\gp))$, too.
Note that the interpretation \irule{I' = \{ b,c,e\}} is not
an answer set for $\gp$. Indeed,
$G_{I'}(\gp) = \{$~\irule{a \Or \tneg b \derives c.} \quad
\irule{\tneg a \Or c.} \quad \irule{e \derives b.} \quad \irule{b.} \quad \irule{c \derives b.} $\}$
and \irule{I'} is not a model for  $pos(G_{I'}(\gp))$, since
the rule \irule{a \Or \tneg b \derives c.} is not satisfied in \irule{I'}.

On the other hand, the interpretation \irule{M} of
Example \ref{ex-3} is an answer set for $\gp$,
since \irule{M} is a minimal model for
$pos(G_{M}(\gp))$.
Moreover, it can be easily realized that
\irule{M} is the only answer set for $\gp$.

Finally, the program $\gp$ of Example \ref{ex-5} admits one
consistent answer set \irule{M_2 = \{ b, \tneg a \}}. Note that
if we replace the strict rule \irule{\tneg a!} by a defeasible
rule \irule{\tneg a.}, $\gp$ admits two answer sets, namely 
\irule{M_1 = \{ a, \tneg b \}} and \irule{M_2 = \{ b, \tneg a \}}.
Asserting \irule{\tneg a}
by a strict rule, prunes the answer set \irule{M_1} stating the
exception (truth of the literal \irule{a}) to this rule. } $\punto$
\end{example}

It is worthwhile noting that if a rule \irule{r} is not satisfied in a
model \irule{M}, then \emph{all} literals in the head of \irule{r}
must be overridden in \irule{M}.

Let $\gp_1$ be the program
\begin{tabbing}
\quad\quad \irule{o2 : o_3} \= \{ \= dummy \kill
\quad\quad \irule{o_3} \> \{ \> \irule{a \Or b.} \quad  \irule{\derives b.} \}\\
\quad\quad \irule{o_2 : o_3} \> \{ \> \irule{\tneg a.} \}
\end{tabbing}
and $\gp_2$
\begin{tabbing}
\quad\quad \irule{o2 : o_1} \= \{ \= dummy \kill
\quad\quad \irule{o_1} \> \{ \> \irule{a.} \quad \irule{\derives b.} \}\\
\quad\quad \irule{o_2 : o_1} \> \{ \> \irule{\tneg a.} \}
\end{tabbing}

Then, \irule{\{\tneg a\}} is not a model for program $\gp_1$,
because the head literal \irule{b} in the head of \irule{a \Or b.} is not overridden
in \irule{M}. If we drop \irule{b} from rule \irule{a \Or b.}, then \irule{\{\tneg a\}} is a model of the
resulting program $\gp_2$.

Observe also that two programs having the same answer sets, as
\irule{o_1} and \irule{o_3} (both have the single answer set
\irule{\{\tneg a\}}), may get different answer sets even if we add the
same object to both of them. Indeed, program $\gp_1$ has no answer
set, while program $\gp_2$ has the answer set \irule{\{\tneg a\}}.

This is not surprising, as a similar phenomenon also arises in normal
logic programming where \irule{P_1 = \{a.\}} and 
\irule{P_2 = \{a \derives \dneg b.\}} have the same answer set \irule{\{a\}},
while \irule{P_1 \cup \{b.\}} and \irule{P_2 \cup \{b.\}} have different
answer sets (\irule{\{a,b\}} and \irule{\{b\}}, respectively).

Finally we show that each answer set of a program $\gp$ is also a
minimal model of $\gp$:

\begin{proposition}\label{is-minimal}
If $M$ is an answer set for $\gp$, then $M$ is
a minimal model of $\gp$.
\end{proposition}

{\noindent
\proof
By contradiction suppose $M'$ is a model for $\gp$ such that
$M' \subset M$.

First we show that $M'$ is a model for $pos(G_M(\gp))$ too,
i.e., every rule in $pos(G_M(\gp))$ is satisfied in $M'$.
Recall that $pos(G_M(\gp))$ is the positive version of the program
obtained by applying the transformation $G_M$ to the program
$ground(\gp)$. Consider a generic rule $r$
of $ground(\gp)$. Since $M'$ is a model
for $\gp$ either (i) $r$ is overridden in $M'$ or (ii)
is satisfied in $M'$.

In case (i), since $M' \subset M$,
from Definition \ref{ov} immediately follows that
$r$ is overridden in $M$ too.
Thus, $r$ does not occur in $G_M(\gp)$
since the transformation $G_M$ removes all rules overridden in $M$.

In case (ii) (i.e., $r$ is satisfied in $M'$), if $r$ is such that
$B(r) \cap M \neq \emptyset$, then the rule is removed by $G_M$.
Otherwise, $r$ is transformed by $G_M$ into a rule $r'$
obtained from $r$ by dropping the NAF part from the body.
Since $r$ is satisfied in $M'$, also $r'$ is satisfied in $M'$.
As a consequence, all the rules of $pos(G_M(\gp))$ are
satisfied in $M'$, that is $M' \subset M$ is a model for $pos(G_M(\gp))$.
Thus, by Definition \ref{answerset}, $M$ is not an answer set
for $\gp$ since it is not a minimal model of $pos(G_M(\gp))$.
The proof is hence concluded.
}

\section{Knowledge Representation with $\DLPI$}\label{KR}

In this section, we present a number of examples which illustrate
how knowledge can be represented using $\DLPI$.
To start, we show the $\DLPI$ encoding of a classical example
of nonmonotonic reasoning.

\begin{example}\label{ex-penguin}
{
Consider the following program $\gp$
with $\O(\gp)$ consisting of three
objects
\irule{bird}, \irule{penguin} and \irule{tweety},  such that
\irule{penguin} is a sub-object of \irule{bird} and
\irule{tweety} is a sub-object of \irule{penguin}:

\begin{tabbing}
\irule{tweety:penguin} \= \{ \= p \kill
\irule{bird}           \> \{ \> \irule{flies.} \  \}
\\
\irule{penguin:bird}   \> \{ \> \irule{\tneg flies!} \  \}
\\
\irule{tweety:penguin} \> \{   \  \}
\end{tabbing}
Unlike in traditional logic programming, our language supports two
types of negation, that is {\em strong negation} and {\em negation
as failure}. Strong negation is useful to express negative pieces
of information under the complete information assumption. Hence, a
negative fact (by strong negation) is true only if it is
explicitly derived from the rules of the program. As a
consequence, the head of rules may contain also such negative
literals and rules can be conflicting on some literals. According
to the inheritance principles, the ordering relationship between
objects can help us to assign different levels of reliability to
the rules, allowing us to solve possible conflicts. For instance,
in our example, the contradicting conclusion {\em tweety both
flies and does not fly} seems to be entailed from the program (as
\irule{tweety} is a \irule{penguin} and \irule{penguin}s are \irule{bird}s, both \irule{flies}
and \irule{\tneg flies} can be derived from the rules of the program).
However, this is not the case. Indeed, the "lower" rule \irule{\tneg
flies.} specified in the object \irule{penguin} is considered
as a sort of refinement to the first general rule, and thus the
meaning of the program is rather clear: {\em tweety does not fly},
as tweety is a penguin. That is, \irule{\tneg flies.} is
preferred to the default rule \irule{flies.} as the hierarchy
explicitly states the specificity of the former. Intuitively,
there is no doubt that \irule{M = \{ \tneg flies \}} is the only
reasonable conclusion. }
$\punto$
\end{example}
The next example, from the field of database authorizations,
combines the use of both weak and strong negation.
\begin{example}\label{ex-security}
{
Consider the following knowledge base representing a set
of security specification about a simple {\em part-of}
hierarchy of objects.
\begin{eqnarray}
\lefteqn{\irule{o_1}} \nonumber\\
& \{ \nonumber\\
&& \irule{authorize(bob) \derives \dneg \ authorize(ann).} \label{rule:auth1}\\
&& \irule{authorize(ann) \Or authorize(tom) \derives \dneg \ \tneg authorize(alice).} \label{rule:auth2}\\
&& \irule{authorize(amy)!} \label{rule:auth3}\\
& \} \nonumber\\
\lefteqn{\irule{o_2:o_1}} \nonumber\\
& \{ \nonumber\\
&& \irule{\tneg authorize(alice)!} \label{rule:auth4}\\
& \} \nonumber\\
\lefteqn{\irule{o_3:o_1}} \nonumber\\
& \{ \nonumber\\
&& \irule{\tneg authorize(bob)!} \label{rule:auth5}\\
& \} \nonumber
\end{eqnarray}

Object \irule{o_2} is  part-of the object \irule{o_1} as well as \irule{o_3}
is part-of \irule{o_1}.
Access authorizations to objects are specified by
rules with head predicate \irule{authorize} and subjects to
which authorizations are granted appear as arguments.
Strong negation is utilized to encode negative authorizations
that represent explicit denials. Negation as failure
is used to specify the
absence of authorization (either positive or negative).
Inheritance implements the automatic propagation
of authorizations from an object to
all its sub-objects.
The overriding mechanism allows us to represent
exceptions: for instance, if an object $o$ inherits a positive
authorization but a denial for the same subject
is specified in $o$, then
the negative authorization prevails on the positive one.
Possible loss of control due to overriding mechanism can be
avoided by using strict rules: strict authorizations cannot
be overridden.

Consider the program $\gp_{o_2} = \{ ( \irule{o_1}, \{ (\ref{rule:auth1}), (\ref{rule:auth2}), (\ref{rule:auth3}) \} ),
( \irule{o_2}, \{ (\ref{rule:auth4}) \} ) \}$ for the object \irule{o_2} on the above knowledge
base. This program defines the access control for the object
\irule{o_2}. Thanks to the inheritance mechanism, authorizations
specified for the object \irule{o_1}, to which \irule{o_2} belongs, are
propagated also to \irule{o_2}. It consists of rules (\ref{rule:auth1}), (\ref{rule:auth2}) and
(\ref{rule:auth3}) (inherited from \irule{o_1}) and (\ref{rule:auth4}). Rule (\ref{rule:auth1}) states that
\irule{bob} is authorized to access object \irule{o_2} provided that no
authorization for \irule{ann} to access \irule{o_2} exists. Rule (\ref{rule:auth2})
authorizes either \irule{ann} or \irule{tom} to access \irule{o_2} provided that no
denial for \irule{alice} to access \irule{o_2} is derived. The strict rule
(\ref{rule:auth3}) grants to \irule{amy} the authorization to access object \irule{o_1}.
Such authorization can be considered "strong", since no exceptions
can be stated to it without producing inconsistency. As a
consequence, all the answer sets of the program contain the
authorization for \irule{amy}. Finally, rule (\ref{rule:auth4}) defines a denial
for \irule{alice} to access object \irule{o_2}. Due to the absence of the
authorization for \irule{ann}, the authorization to \irule{bob} of accessing
the object \irule{o_2} is derived (by rule (\ref{rule:auth1})). Further, the explicit
denial to access the object \irule{o_2} for \irule{alice} (rule (\ref{rule:auth4})) allows
to derive neither authorization for \irule{ann} nor for \irule{tom} (by rule
(\ref{rule:auth2})). Hence, the only answer set of this program is \irule{\{
authorize(bob), \tneg authorize(alice), authorize(amy) \}}.

Consider now the program
$\gp_{o_3} = \{ ( \irule{o_1}, \{ (\ref{rule:auth1}), (\ref{rule:auth2}) \} ), ( \irule{o_3}, \{ (\ref{rule:auth5}) \} ) \}$
for the object \irule{o_3}.
Rule (\ref{rule:auth5}) defines a denial for \irule{bob} to access
object \irule{o_3}.
The authorization for \irule{bob} (defined by rule (\ref{rule:auth1})) is no longer derived.
Indeed, even if rule (\ref{rule:auth1}) allows to derive
such an authorization due to the absence of authorizations for
\irule{ann}, it is overridden by the explicit denial (rule (\ref{rule:auth5})) defined
in the object \irule{o_3} (i.e., at a more specific level).
The body of rule (\ref{rule:auth2}) inherited from \irule{o_1} is true for this program
since no denial for alice can be derived, and it entails a mutual exclusive access to object
\irule{o_3} for \irule{ann} and \irule{tom} (note that no other head contains
\irule{authorize(ann)} or \irule{authorize(bob)}).
The program $\gp_{o_3}$ admits two answer sets, namely
\{\irule{authorize(ann)}, \irule{\tneg authorize(bob)}, \irule{authorize(amy)}\} and
\{\irule{authorize(tom)}, \irule{\tneg authorize(bob)}, \irule{authorize(amy)}\} representing
two alternative authorization sets to grant
the access to the object \irule{o_3}.
}
$\punto$
\end{example}

\subsection*{Solving the Frame Problem}

The frame problem has first been addressed by McCarthy and Hayes
\cite{mcca-haye-69}, and in the meantime a lot of research has been
conducted to overcome it (see e.g.\ \cite{shan-97} for a survey).

In short, the frame problem arises in planning, when actions and
fluents are specified: An action affects some of the fluents, but all
unrelated fluents should remain as they are. In most formulations
using classical logic, one must specify for every pair of
actions and unrelated fluents that the fluent remains unchanged.
Clearly this is an undesirable overhead, since with $n$ actions and
$m$ fluents, $n \times m$ clauses would be needed.

Instead, it would be nice to be able to specify for each fluent that it
{\em ``normally remains valid''} and that only actions which explicitly
entail the contrary can change them.

Indeed, this goal can be achieved in a very elegant way using $\DLPI$:
One object contains the rules which specify {\em inertia} (the fact
that fluents normally do not change). Another object inherits from it
and specifies the actions and the effects of actions --- in this way a
very natural, straightforward and effective representation is
achieved, which avoids the frame problem.

\begin{example}\label{ex-yaleshooting}
As an example we show how the famous Yale Shooting Problem, which is due to
Hanks and McDermott \cite{hank-mcde-87}, can be represented and
solved with $\DLPI$:

The scenario involves an individual (or in a less violent version a
turkey), who can be shot with a gun.  There are two fluents, \irule{alive}
and \irule{loaded}, which intuitively mean that the individual is alive and
that the gun is loaded, respectively. There are three actions,
\irule{load}, \irule{wait} and \irule{shoot}. Loading has the effect that the gun is
loaded afterwards, shooting with the loaded gun has the effect that
the individual is no longer alive afterwards (and also that the gun is
unloaded, but this not really important), and waiting has no
effects.

The problem involves {\em temporal projection}: It is known that initially the individual is alive, and
that first the gun is loaded, and after waiting, the gun is shot
with. The question is: Which fluents hold after these actions and
between them?

In our encoding, the \irule{inertia} object contains the defaults for the
fluents, the \irule{domain} object additionally specifies the effects of
actions, while the \irule{yale} object encodes the problem instance.

For the time framework we use the \dlv bounded integer built-ins: The
upper bound \irule{n} of positive integers is specified by either adding the
fact \irule{\#maxint = n.} to the program or by passing the option
\irule{-N=n} on the commandline (this overrides any \irule{\#maxint =
n.} statement). It is then possible to use the built-in constant
\irule{\#maxint}, which evaluates to the specified upper bound, and
several built-in predicates, of which in this paper we just use
\irule{\#succ(N,N1)} , which holds if \irule{N1} is the successor of
\irule{N} and \irule{N1 \leq \#maxint}. For additional \dlv built-in
predicates, consult the \dlv homepage \cite{dlv-web}.

\begin{eqnarray}
\lefteqn{\irule{inertia}} \nonumber\\
& \{ \nonumber\\
&& \irule{alive(T1) \derives alive(T), \#succ(T,T1).}\\
&& \irule{\tneg alive(T1) \derives \tneg alive(T), \#succ(T,T1).}\\
&& \irule{loaded(T1) \derives loaded(T), \#succ(T,T1).}\\
&& \irule{\tneg loaded(T1) \derives \tneg loaded(T), \#succ(T,T1).}\\
& \} \nonumber\\
\lefteqn{\irule{domain : inertia}} \nonumber\\
& \{\\
&& \irule{loaded(T1) \derives load(T), \#succ(T,T1)!} \\
&& \irule{\tneg loaded(T1) \derives shoot(T), loaded(T), \#succ(T,T1)!} \\
&& \irule{\tneg alive(T1) \derives shoot(T), loaded(T), \#succ(T,T1)!} \\
& \} \nonumber\\
\lefteqn{\irule{yale : domain}} \nonumber\\
& \{\\
&& \irule{load(0)! \quad wait(1)! \quad shoot(2)! \quad alive(0)!} \\
& \} \nonumber
\end{eqnarray}

The only answer set for this program (and \irule{\#maxint = 3}) contains, besides the facts of the \irule{yale}
object, \irule{loaded(1)}, \irule{loaded(2)},
\irule{alive(0)}, \irule{alive(1)}, \irule{alive(2)} and
\irule{\tneg loaded(3)}, \irule{\tneg alive(3)}.
That is, the individual is alive until the shoot action is
taken, and no longer alive afterwards, and the gun is loaded between
loading and shooting.
$\punto$
\end{example}

We want to point out that this formalism is equally suited for solving
problems which involve finding a plan (i.e.\ a sequence of actions)
rather than doing temporal projection (determining the effects of a given plan) as in the Yale
Shooting Problem: You have to add a rule \irule{action(T) \Or \tneg action(T)
\derives \#succ(T,T1).} for every action, and you have to specify the
goal state by a query, e.g. \irule{\tneg alive(3), \tneg
loaded(3)?} A query is a \dlv language feature which (for this example) is equivalent to
the rules \irule{h \derives \tneg alive(3), \tneg loaded(3).} and 
\irule{i \derives \dneg \ h, \dneg \ i.}, meaning that only answer sets containing
\irule{\tneg alive(3)} and \irule{\tneg loaded(3)} should be considered.

Below you find a classical plan-finding example: The blocksworld domain and the
Sussman anomaly as a concrete problem.

\begin{example}
In \cite{erde-99}, several planning problems, including the
blocksworld problems, are encoded using disjunctive datalog.

In general, planning problems can be effectively specified using
action languages (e.g.\
\cite{gelf-lifs-93,dung-93,giun-lifs-98,lifs-99}).  Then, a
translation from these languages to another language (in our case
$\DLPI$) is applied.

We omit the step of describing an action
language and the associated translation, and directly show the
encoding of an example planning domain in disjunctive datalog. This
encoding is rather different from the one presented in \cite{erde-99}.

The objects in the blocksworld are one \irule{table} and an arbitrary number
of labeled cubic \irule{block}s. Together, they are referred to as
\irule{loc}ations.

The state of the blocksworld at a particular time can be fully
specified by the fluent \irule{on(B,L,T)}, which specifies that
block \irule{B} resides on location \irule{L} at time \irule{T}.

So, first we state in the object \irule{bw\_inertia} that the fluent
\irule{on} is inertial.
\begin{eqnarray}
\lefteqn{\irule{bw\_inertia}} \nonumber\\
& \{ \nonumber\\
&& \irule{on(B,L,T1) \derives on(B,L,T), \#succ(T,T1).}\\
& \} \nonumber
\end{eqnarray}

We continue to define the blocksworld domain in the object \irule{bw\_domain}, which inherits from the inertia object:
\begin{eqnarray}{}
\lefteqn{\irule{bw\_domain : bw\_inertia}} \nonumber\\
& \{ \nonumber\\
&& \irule{move(B,L,T) \Or \tneg move(B,L,T) \derives block(B), loc(L), \#succ(T,T1)!} \label{rule:bw:guess}\\
&&                               \irule{on(B,L,T1) \derives move(B,L,T), \#succ(T,T1)!} \label{rule:bw:on_effect}\\
&&                               \irule{\tneg on(B,L,T1) \derives move(B,L1,T), on(B,L,T), \#succ(T,T1)!} \label{rule:bw:not_on_effect}\\
&&                               \irule{\derives move(B,L,T), on(B1,B,T).} \label{rule:bw:constraint_clear}\\
&&                               \irule{\derives move(B,B1,T), on(B2,B1,T), block(B1).} \label{rule:bw:constraint_clear2}\\
&&                               \irule{\derives move(B,B,T).} \label{rule:bw:constraint_notonitself}\\
&&                               \irule{\derives move(B,L,T), move(B1,L1,T), B<>B1.} \label{rule:bw:constraint_uniquesource}\\
&&                               \irule{\derives move(B,L,T), move(B1,L1,T), L<>L1.} \label{rule:bw:constraint_uniquedestination}\\
&&                               \irule{loc(table)!} \\
&&                               \irule{loc(B) \derives block(B)!}\\
& \} \nonumber
\end{eqnarray}

There is one action, which is moving a block from one location to
another location. A move is started at one point in time, and it is
completed before the next time. Rule (\ref{rule:bw:guess}) expresses that at any time \irule{T}, the action of
moving a block \irule{B} to location \irule{L} may be initiated
(\irule{move(B,L,T)}) or not (\irule{\tneg move(B,L,T)}).

Rules (\ref{rule:bw:on_effect}) and (\ref{rule:bw:not_on_effect}) specify the
effects of the move action: The
moved block is at the target location at the next time, and no longer
on the source location.

(\ref{rule:bw:constraint_clear}) -- (\ref{rule:bw:constraint_uniquedestination})
are constraints, and their semantics is that in any
answer set the conjunction of their literals must not be true.%
\footnote{\label{foot:constraints}We use constraints for clarity, but they can be eliminated
by rewriting \irule{\derives B.} to \irule{p \derives B, \dneg \ p.},
where \irule{p} is a new symbol which does not appear anywhere else
in the program.}
Their respective meanings are:
(\ref{rule:bw:constraint_clear}): A moved block must be clear.
(\ref{rule:bw:constraint_clear2}): The target of a move must be clear if it is a block (the table may hold an arbitrary number of blocks).
(\ref{rule:bw:constraint_notonitself}) A block may not be on itself.
(\ref{rule:bw:constraint_uniquesource}) and (\ref{rule:bw:constraint_uniquedestination}): No two move actions may be performed at the same time.

The timesteps are again represented by \dlv's integer built-in
predicates and constants.

\begin{figure}
\begin{center}
\epsfig{file=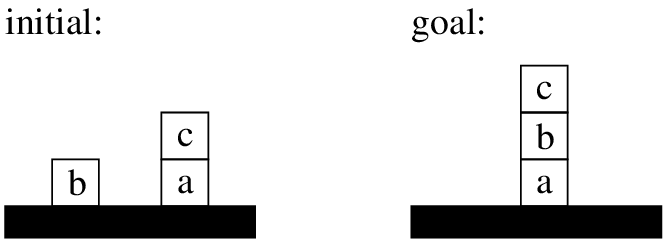}
\end{center}
\caption{The Sussman Anomaly}
\label{sussman}
\end{figure}

What is left is the concrete problem instance, in our case the
so-called Sussman Anomaly (see Figure \ref{sussman}):
\begin{eqnarray}
\lefteqn{\irule{sussman : bw\_domain}} \nonumber\\
& \{ \nonumber\\
&& \irule{block(a)!} \  \quad \irule{block(b)!} \  \quad \irule{block(c)!} \label{rule:bwi:blocks}\\
&& \irule{on(b, table, 0)!} \  \quad \irule{on(c, a, 0)!} \  \quad \irule{on(a, table, 0)!} \label{rule:bwi:initial_situation}\\
& \} \nonumber\\
\lefteqn{\irule{on(c, b, \#maxint),on(b, a, \#maxint), on(a, table, \#maxint) ?}} \label{rule:bwi:query}
\end{eqnarray}

Since different problem
instances may involve different numbers of blocks, the blocks are
defined as facts (\ref{rule:bwi:blocks}) together with the problem instance.%
\footnote{Note that usually the instance will be separated from
the domain definition.}

We give the initial situation by facts (\ref{rule:bwi:initial_situation}), while the goal situation is
specified by query (\ref{rule:bwi:query}). This query enforces that only those answer sets
are computed, in which the conjunction of the query literals is true.

\end{example}

\section{Computational Complexity}
\label{sec:complexity}
\def\parag{{\vspace{0.15in} \noindent}}

As for the classical nonmonotonic formalisms
\cite{mare-trus-91,mare-trus-90,reit-80},
two important decision problems, corresponding to two different
reasoning tasks, arise in $\DLPI$:
\\
{\em (Brave Reasoning)}
Given a $\DLPI$ program $\gp$ and a ground literal $L$,
decide whether there exists an answer set $M$ for $\gp$
such that $L$ is true w.r.t. $M$.
\\
{\em (Cautious Reasoning)}
Given a $\DLPI$ program $\gp$ and a ground literal $L$,
decide whether $L$ is true in all answer sets for $\gp$.

We next prove that the complexity of reasoning
in $\DLPI$ is exactly
the same as in traditional disjunctive logic programming.
That is, inheritance comes for free, as the addition of inheritance
does not cause any computational overhead.
We consider the propositional case, i.e., we consider ground
$\DLPI$ programs.

\begin{lemma}\label{recognition}
Given a ground $\DLPI$ program $\gp$ and an interpretation
$M$ for $\gp$,
deciding whether $M$ is an answer set
for $\gp$ is in $\CONP$.
\end{lemma}

\begin{proof}
We check in $\NP$ that $M$ \textbf{is not} an answer set of $\gp$
as follows.
Guess a subset $I$ of $M$, and verify that:
(1) $M$ is not a model for $pos(G_M(\gp))$, or
(2) $I$ is a model for $pos(G_M(\gp))$ and $I\subset M$.
The construction of $pos(G_M(\gp))$ (see Definition \ref{reduction})
is feasible in polynomial time, and the tasks (1) and (2) are
clearly tractable.
Thus, deciding whether $M$ is not an answer set for $\gp$ is in $\NP$,
and, consequently, deciding whether $M$ is an answer set
for $\gp$ is in $\CONP$.
\end{proof}

\begin{theorem}\label{brave}
Brave Reasoning on $\DLPI$ programs is $\SigmaP{2}$-complete.
\end{theorem}

\begin{proof}
Given a ground $\DLPI$ program $\gp$ and a ground literal
$L$, we verify that $L$ is a brave consequence of $\gp$ as follows.
Guess a set $M \subseteq \BP$ of ground literals, check that
(1) $M$ is an answer set for $\gp$, and (2) $L$ is true w.r.t. $M$.
Task (2) is clearly polynomial; while (1) is in $\CONP$,
by virtue of Lemma \ref{recognition}.
The problem therefore lies in $\SigmaP{2}$.

$\SigmaP{2}$-hardness follows from Theorem
\ref{the:answerset} and the results in \cite{eite-gott-95,eite-etal-97f}.
\end{proof}

\begin{theorem}\label{cautious}
Cautious Reasoning on $\DLPI$ programs is $\PiP{2}$-complete.
\end{theorem}

\begin{proof}
Given a ground $\DLPI$ program $\gp$ and a ground literal
$L$, we verify that $L$ is not a cautious consequence of $\gp$ as follows.
Guess a set $M \subseteq \BP$ of ground literals, check that
(1) $M$ is an answer set for $\gp$, and (2) $L$ is not true w.r.t. $M$.
Task (2) is clearly polynomial; while (1) is in $\CONP$,
by virtue of Lemma \ref{recognition}.
Therefore, the complement of cautious reasoning is in $\SigmaP{2}$,
and cautious reasoning is in $\PiP{2}$.

$\PiP{2}$-hardness follows from Theorem
\ref{the:answerset} and the results in \cite{eite-gott-95,eite-etal-97f}.
\end{proof}

\section{Related Work}\label{sec:relatedwork}

\subsection{Answer Set Semantics}

Answer Set Semantics, proposed by Gelfond and Lifschitz
in \cite{gelf-lifs-91}, is the most widely acknowledged semantics
for disjunctive logic programs with strong negation.
For this reason, while defining the semantics of our language,
we took care of ensuring full agreement with Answer Set Semantics
(on inheritance-free programs).

\begin{theorem}\label{the:answerset}
Let $\gp$ be a $\DLPI$ program consisting of a single object
$o=\tuple{oid(o),\Sigma(o)}$.%
\footnote{
On inheritance-free programs, there is no difference
between strict and defeasible rules.
Therefore, without loss of generality
we assume that rules are of only one type here.
This allows us to drop the symbol ('.' or '!') at the end of the rules
of single object programs.}
Then, $M$ is an answer set of $\gp$ if and only if it is a consistent
answer set of $\Sigma(o)$ (as defined in \cite{gelf-lifs-91}).
\end{theorem}

\begin{proof}
First we show that $G_M(\gp)$ is equal to $\Sigma(o)^M$ (as defined in
\cite{gelf-lifs-91}):

Deletion rule (1) of Definition \ref{reduction} never applies,
since for every literal L and any two rules $r_1$, $r_2$ $\in
ground(\gp)$, $obj\_of(r_1) \not < obj\_of(r_2)$ holds, thus
violating condition (1) in Definition \ref{ov} and therefore no
rule can be overridden. It is evident that the deletion rules (2)
and (3) of Definition \ref{reduction} are equal to deletion rules
(i) and (ii) of the definition of $\Pi^S$ in \S 7 in
\cite{gelf-lifs-91}, respectively. The first ones delete rules, where
some NAF literal is contained in $M$, while the second ones delete
all NAF literals of the remaining rules.

Next, we show that the criteria for a consistent set $M$ of
literals being an answer set of a positive (i.e. NAF free)
program (as in \cite{gelf-lifs-91}) is equal to the notion of
satisfaction:

Since the set is consistent, condition (ii) in \S 7 of \cite{gelf-lifs-91}
does not apply. Condition (i) says: $L_{k+1}, \dots, L_{m} \in M$ (the
body is true) implies that the head is true. This is logically equivalent to ``The body is not true or the head is true'', which is the definition of rule satisfaction.

In total we have that the minimal models of $pos(G_M(\gp))$ are equal
to the consistent answer sets of $\Sigma(o)^M$, since answer sets are
minimal by definition.

Additionally, we require in Definition \ref{answerset} that $M$ is
also a model of $\gp$, while in \cite{gelf-lifs-91} there is no such
requirement. However, all minimal models of $pos(G_M(\gp))$ are also
models of $\gp$: All rules in $G_M(\gp)$ are satisfied, and only the
deletion rules (2) and (3) of Definition \ref{reduction} have been
applied (as shown above). So, for any rule $r$, which has been deleted
by (2), some literal in $Body^-(r)$ is in $M$, so $r$'s body is not
true, and thus $r$ is satisfied in $M$. If a rule $r$, which has been
transformed by (3), is satisfied without $Body^-(r)$, then either
$Head(r)$ is true or $Body^+(r)$ is not true, so adding any NAF part
to it does not change its satisfaction status.
\end{proof}

Theorem \ref{the:answerset} shows that the set of rules contained in a
single object of a $\DLPI$ program has precisely the same answer sets
(according to the definition in \cite{gelf-lifs-91}) as the single object
program (according to Definition~\ref{answerset}).

For a $\DLPI$ program
$\gp$ consisting of more than one object, the answer sets (as defined in
\cite{gelf-lifs-91}) of the collection of all rules in $\gp$ in general do not coincide with the answer sets of $\gp$.

For instance the program
\begin{tabbing}
\irule{o1:o} \= \{ \= p \kill
\irule{o}           \> \{ \> \irule{p.} \  \}
\\
\irule{o1:o}   \> \{ \> \irule{\tneg p.} \  \}
\end{tabbing}
has the answer set \irule{\{\tneg p \}}, while the disjunctive logic program
\irule{\{p.\ \tneg p.\}} does not have a consistent answer set.

Nevertheless, in Section~\ref{sec:implementation_I} we will show that
each $\DLPI$ program $\gp$ can be translated into a disjunctive logic
program $\gp'$, the semantics of which is equivalent to the semantics
of $\gp$. However, this translation requires the addition of a number
of extra predicates.

\subsection{Disjunctive Ordered Logic}\label{DOL}

Disjunctive Ordered Logic ($\DOL$) is an extension of
Disjunctive Logic Programming with strong negation and inheritance
(without default negation)
proposed in \cite{bucc-etal-98a,bucc-etal-99c}.
The $\DLPI$ language incorporates some ideas taken from $\DOL$.
However, the two languages are very different in several respects.
Most importantly, unlike with $\DLPI$,
even if a program belongs to the common fragment of $\DOL$
and of the language of \cite{gelf-lifs-91}
(i.e., it contains neither inheritance nor default negation),
$\DOL$ semantics is completely different from Answer Set Semantics,
because of a different way of handling contradictions.%
\footnote{Actually, this was a main motivation for the authors
to look for a different language.}
In short, we observe the following differences between $\DOL$ and
$\DLPI$:
\begin{itemize}
\item
$\DOL$ does not include default negation $\dneg$,
while $\DLPI$ does.
\item
$\DOL$ and $\DLPI$ have different semantics
on the common fragment.
Consider a program $\gp$ consisting of a single object
$o=\tuple{oid(o),\Sigma(o)}$, where
$\Sigma(o)=\{\irule{p.}\quad\irule{\tneg p.} \}$.
Then, according to $\DOL$, the semantics of $\gp$
is given by two models, namely, \irule{\{p\}} and \irule{\{\tneg p\}}.
On the contrary, $\gp$ has no answer set according to
$\DLPI$ semantics.
\item
$\DLPI$ generalizes (consistent) Answer Set Semantics to
disjunctive logic programs with inheritance, while $\DOL$ does not.
\end{itemize}

\subsection{Prioritized Logic Programs}

$\DLPI$ can be also seen as an attempt to handle priorities in
disjunctive logic programs
(the lower the object in the inheritance hierarchy,
the higher the priority of its rules).

There are several works on preference handling in logic programming
\cite{delg-etal-00a,brew-eite-98,gelf-son-97,nute-94,kowa-sadr-90,prad-mink-96,saka-inou-96}.
However, we are aware of only one previous work on priorities
in \textbf{disjunctive} programs, namely, the
paper by Sakama and Inoue \cite{saka-inou-96}.
This interesting work can be seen as an extension of
Answer Set Semantics to deal with priorities.
Comparing the two approaches under the perspective of priority handling,
we observe the following:
\begin{itemize}
\item
On priority-free programs, the two languages yield essentially
the same semantics, as they generalize Answer Set Semantics and
Consistent Answer Set Semantics, respectively.
\item
In \cite{saka-inou-96}, priorities are defined among 
\textbf{literals}, while priorities concern program
\textbf{rules} in $\DLPI$.
\item
The different kind of priorities (on rules vs.\ literals)
and the way how they are dealt with in the two approaches
imply different complexity in the respective reasoning tasks.
Indeed, from the simulation of abductive reasoning
in the language of \cite{saka-inou-96}, and the complexity
results on abduction reported in \cite{eite-etal-97k},
it follows that brave reasoning is $\SigmaP{3}$-complete
for the language of \cite{saka-inou-96}.
On the contrary, brave reasoning is ``only''
$\SigmaP{2}$-complete in $\DLPI$%
\footnote{We refer to the complexity in the propositional case here.}.
\end{itemize}

\cite{delg-etal-00a} deals with nondisjunctive programs, but the
authors note that their semantics-defining transformation ``is also
applicable to disjunctive logic programs''. In this formalism, the
preference relation is defined by regular atoms (with a set of
constants representing the rules), allowing the definition of
dynamic preferences. However, the semantics of the preferences is
based on the order of rule application (or defeating) and thus seems to be
quite different from our approach.

A comparative analysis of the various approaches to the treatment
of preferences in ($\vee$-free)
logic programming has been carried out
in \cite{brew-eite-98}.

\subsection{Inheritance Networks}

From a different perspective, the objects of a
$\DLPI$ program can also be seen as the nodes of an inheritance network.

We next show that $\DLPI$ satisfies the basic semantic principles
which are required for inheritance networks in \cite{tour-86}.

\cite{tour-86} constitutes a fundamental attempt to present a formal
mathematical theory of multiple inheritance with exceptions.
The starting point of this work is the consideration
that an intuitively acceptable semantics for inheritance
must satisfy two basic requirements:
\begin{enumerate}
\item
Being able to reason with redundant statements, and
\item
not making unjustified choices in ambiguous situations.
\end{enumerate}

Touretzky illustrates this intuition by means of two basic examples.

The former requirement is presented by means of the {\em Royal
Elephant} example, in which we have the following knowledge:
``Elephants are gray.'', ``Royal elephants are elephants.'', ``Royal
elephants are not gray.'', ``Clyde is a royal elephant.'', ``Clyde is
an elephant.''

The last statement is clearly redundant; however, since it is
consistent with the others there is no reason
to rule it out.
Touretzky shows that an intuitive semantics should be able
to recognize that Clyde is not gray, while many systems fail in
this task.

Touretzky's second principle is shown by the {\em Nixon diamond} example, in which the following is known:
``Republicans are not pacifists.'',
``Quakers are pacifists.'',
``Nixon is both a Republican and a quaker.''

According to our approach, he claims that a good semantics should draw
no conclusion about the question whether Nixon is a $pacifist$.

The proposed solution for the problems above
is based on a topological relation,
called {\em inferential distance ordering}, stating that an individual
$A$ is "nearer" to $B$ than to $C$ iff $A$ has an inference
path {\em through} $B$ to $C$.
If $A$ is "nearer" to $B$ than to $C$, then as far as $A$ is concerned,
information coming from $B$ must be preferred w.r.t. information
coming from $C$.
Therefore, since Clyde is "nearer" to being a royal elephant than to
being an elephant, he states that Clyde is not gray.
On the contrary no conclusion is taken on Nixon, as there is not
any relationship between quaker and republican.

The semantics of $\DLPI$ fully agrees with the intuition underlying the
{\em inferential distance ordering}.

\begin{example}
Let us represent the Royal Elephant example in our
framework:
\vspace{-0.2cm}
\[
\begin{array}{ll}
\irule{elephant} & \{ \irule{gray.} \}\\
\irule{royal\_elephant: elephant} & \{ \irule{\tneg gray.} \}\\
\irule{clyde:elephant,\ royal\_elephant} & \{\ \}
\end{array}
\vspace{-0.1cm}
\]
The only answer set of the above
$\DLPI$ program is \irule{\{\tneg gray\}}.

The Nixon Diamond example can be expressed in our language as follows:
\vspace{-0.2cm}
\[
\begin{array}{ll}
\irule{republican} & \{ \irule{\tneg pacifist.} \}\\
\irule{quaker} & \{ \irule{pacifist.} \}\\
\irule{nixon : republican, quaker} & \{\ \}
\end{array}
\vspace{-0.1cm}
\]
This $\DLPI$ program has no
answer set, and therefore no conclusion is drawn.
\end{example}

\subsection{Updates in Logic Programs}

The definition of the semantics of updates in logic programs is
another topic where $\DLPI$ could potentially be applied.
Roughly, a simple formulation of the problem is the following:
Given a ($\vee$-free) logic program $P$ and
a sequence $U_1,\cdots,U_n$ of successive updates
(insertion/deletion of ground atoms),
determine what is or is not true in the end.
Expressing the insertion (deletion) of an atom $A$ by the rule
$A\derives$ ($\tneg A\derives$),
we can represent this problem by a $\DLPI$ knowledge base
$\{\tuple{t_0,P},\tuple{t_1,\{U_1\}},\cdots,$ $\tuple{t_n,\{U_n\}}\}$
($t_i$ intuitively represents the instant of time when the update
$U_i$ has been executed), where $t_n<\cdots <t_0$.%
\footnote{In this context, $<$ should be interpreted as ``more recent''.}
The answer sets of the program for $t_k$ can be taken as the
semantics of the execution of $U_1,\cdots,U_k$ on $P$.
For instance, given the logic program \irule{P=\{a\derives b, \dneg\; c\}} and the updates
\irule{U_1=\{b.\}}, \irule{U_2=\{c.\}},
\irule{U_3=\{\tneg b. \}}, we build the $\DLPI$ program
\vspace{-0.2cm}
\begin{tabbing}
$\quad \irule{t_0} \quad\quad\quad$ \= $ \{\;\irule{a\derives b, c, \dneg\; d.}\;\}$
\\
$\quad \irule{t_1:t_0}$ \> $\{\; \irule{b.} \;\}$
\\
$\quad \irule{t_2:t_1}$ \> $\{\; \irule{c.} \;\}$
\\
$\quad \irule{t_3:t_2}$ \> $\{\; \irule{\tneg b.} \;\}$.
\vspace{-0.1cm}
\end{tabbing}

The answer set \irule{\{a,b,c \}} of the program for \irule{t_2} gives the semantics of the
execution of \irule{U_1} and \irule{U_2} on \irule{P}; while the answer set \irule{\{c\}}
of the program for \irule{t_3} expresses the semantics of the execution
of \irule{U_1}, \irule{U_2} and \irule{U_3} on \irule{P} in the given order.

The semantics of updates obtained in this way is very similar
to the approach adopted for the ULL language in \cite{leon-etal-95b}.
Further investigations are needed on this topic to see whether
$\DLPI$ can represent update problems in more general settings
like those treated in \cite{mare-trus-94} and in \cite{alfe-etal-98b}.
A comparative analysis of various approaches to updating logic programs
is being carried out in \cite{eite-etal-2000-jelia}.
Preliminaries results of this work show that,
under suitable syntactic conditions,
$\DLPI$ supports a nice ``iterative'' property
for updates, which is missed in other formalisms.

\section{Implementation Issues}\label{sec:implementation}
\subsection{From $\DLPI$ to Plain DLP}\label{sec:implementation_I}

In this section we show how a $\DLPI$ program can be translated into
an equivalent plain disjunctive logic program (with strong negation,
but without inheritance).  The translation allows us to exploit
existing disjunctive logic programming (DLP) systems for the implementation
of $\DLPI$.

\noindent
{\bf Notation.}
\begin{enumerate}
\item
Let $\gp$ be the input $\DLPI$ program.
\item
We denote  a literal by $\phi (\bar X)$,
where $\bar X$ is the tuple of the literal's arguments, and
$\phi$ represents an {\em adorned predicate}, that is either
a predicate symbol $p$
or a strongly negated predicated symbol $\tneg p$.
Two adorned predicates are {\em complementary} if one is the negation
of the other (e.g., $q$ and $\tneg q$ are complementary).
$\tneg.\phi$ denotes the complementary adorned predicate of the adorned
predicate $\phi$.
\item
An adorned predicate $\phi$ is
{\em conflicting} if both $\phi(\bar{X})$ and $\tneg.\phi(\bar Y)$ occur in the heads of rules in $\gp$.
\item
Given an object $o$ in $\gp$,
and a head literal $\phi(\bar X)$ of a defeasible rule in $\Sigma(o)$,
we say that $\phi$ is {\em threatened in o} if a literal $\tneg.\phi(\bar Y)$
occurs in the head of a rule in $\Sigma(o')$
where $o' < o$.
A defeasible rule $r$ in $\Sigma(o)$ is {\em threatened in o} if
all its head literals are threatened in $o$.
\end{enumerate}
The rewriting algorithm translating $\DLPI$ programs
in plain disjunctive logic programs with constraints\footnote{Again we use constraints for clarity, see footnote \ref{foot:constraints} on page \pageref{foot:constraints}}
is shown in Figure \ref{alg1}.

\begin{figure}
\begin{algorithmic}
\STATE \hspace{-0.3 cm} {\bf ALGORITHM}
\STATE \hspace{-0.3 cm} {\bf INPUT}: a $\DLPI$-program $\gp$
\STATE \hspace{-0.3 cm}
{\bf OUTPUT}: a plain disjunctive logic program with constraints $DLP(\gp)$
\end{algorithmic}
\begin{algorithmic}[1]
\STATE $DLP(\gp) \Leftarrow \{ prec'(o,o_1) \derives \ | \ o < o_1 \}$
\FOR{each object $o \in \O(\gp)$}
\FOR{each threatened adorned predicate $\phi$ in $o$}
\STATE Add the following rule to $DLP(\gp)$:
\STATE \quad $ovr'(\phi,o,X_1,\cdots,X_n) \derives \tneg . \phi'(X,X_1,\cdots,X_n), prec'(X,o)$
\STATE where $n$ is the arity of $\phi$ and $X,X_1,\cdots,X_n$ are distinct variables.
\ENDFOR
\FOR{each rule $r$ in $\Sigma(o)$, say
$\phi_1(\bar X_1) \Or  \cdots \Or \phi_n(\bar X_n) \derives BODY$,}
\IF{$r$ is threatened}
\STATE Add the following two rules to $DLP(\gp)$:
\STATE \quad $\phi'_1( o,\bar X_1) \Or \cdots \Or \phi'_n(o,\bar X_n) \derives BODY, \ \dneg \ ovr'(r,o,\bar X_1, ... , \bar X_n)$
\STATE \quad $ovr'(r,o,\bar X_1, ... , \bar X_n) \derives ovr'({\phi_1},o,\bar X_1), ... , ovr'({\phi_n},o, \bar X_n)$
\ELSE
\STATE Add the following rule to $DLP(\gp)$:
\STATE \quad $\phi'_1( o,\bar X_1) \Or \cdots \Or \phi'_n(o,\bar X_n) \derives BODY$
\ENDIF
\ENDFOR
\ENDFOR
\FOR{each adorned predicate $\phi$ appearing in $\gp$}
\STATE Add the following rule to $DLP(\gp)$:
\STATE \quad $\phi(X_1,\cdots,X_n) \derives \phi'(X_0,X_1,\cdots,X_n)$
\STATE where $n$ is the arity of $\phi$ and $X_0,\cdots,X_n$ are distinct variables.
\ENDFOR
\FOR{each conflicting adorned predicate $\phi$ appearing in $\gp$}
\STATE Add the following constraint to $DLP(\gp)$:
\STATE \quad $\derives \phi(X_1,\cdots,X_n), \tneg . \phi(X_1,\cdots,X_n)$
\STATE where $n$ is the arity of $\phi$ and $X_1,\cdots,X_n$ are distinct variables.
\ENDFOR
\end{algorithmic}
\caption{A Rewriting Algorithm}
\label{alg1}
\end{figure}

An informal description of how the algorithm proceeds is
the following:
\begin{itemize}
\item
$DLP(\gp)$ is initialized to a set of facts with head
predicate $prec'$ representing the partial ordering among objects
(statement 1).

\item
Then, for each object $o$ in $\O(\gp)$:
\begin{itemize}
\item
For each threatened literal $\phi(\bar X)$ appearing in $o$,
rules defining when the literal is overridden are added
(statements 3--7).
\item
For each rule $r$ belonging to $o$:
\begin{enumerate}
\item
If $r$ is threatened, then the rule is rewritten, such that the head
literals include information about the object in which they have been
derived, and the body includes a literal which satisfies the rule
if it is overridden.
In addition, a rule is added which encodes when the rule is overridden (statements 9--12).

\item
Otherwise (i.e., if $r$ is not threatened) just the rule head is rewritten as described above,
since these rules cannot be overridden (statements 13--15).
\end{enumerate}
\end{itemize}
\item
For all adorned predicates in the program, we add a rule which states
that an atom with this predicate holds, no matter in which object it
has been derived (statements 19--23). The information in which object
an atom has been derived is only needed for determination of
overriding.

\item
Finally, statements 24--28 add a constraint for each adorned predicate, which prevents the generation of inconsistent sets of literals.
\end{itemize}

$DLP(\gp)$ is referred to as the {\em DLP version} of the program $\gp$.%
\footnote{$DLP(\gp)$ is a function-free disjunctive logic program.
  Allowing functions could make the algorithm notation more compact,
  but would not give any computational benefit.}

We now give an example to show how the translation works:

\begin{example}\label{ex-translation}
{
The datalog version $DLP(\gp)$ of the program
$\gp$ of Example \ref{ex-1} is:

\noindent
$\begin{array}{ll}
\bf (1) & \bf rules \ expliciting \ partial \ order \ among \ objects:\\
& \irule{prec'(o_2,o_1).}\\
\end{array}$\\
$\begin{array}{ll}
{\bf (2)} & {\bf rules \ for \ threatened \ adorned \ predicates \ in} \ o_1:\\
& \irule{ovr'(a,o_1) \derives \tneg a'(X), prec'(X,o_1).}\\
& \irule{ovr'(\tneg b, o_1)  \derives b'(X), prec'(X,o_1).}\\
\end{array}$\\
$\begin{array}{ll}
{\bf (3)} & {\bf rewriting \ of \ rules \ in} \ o_1:\\
& \irule{a'(o_1) \Or \tneg b'(o_1) \derives  c, \dneg \ d, \dneg \ ovr'(r_1,o_1).} \\
& \irule{ovr'(r_1,o_1)  \derives ovr'(a,o_1), ovr'(\tneg b,o_1).}\\
& \irule{e'(o_1) \derives b.}\\
\end{array}$\\
$\begin{array}{ll}
{\bf (4)} & {\bf rewriting \ of \ rules \ in} \ o_2:\\
& \irule{\tneg a'(o_2) \Or  c'(o_2).}\\
& \irule{b'(o_2).}\\
& \irule{c'(o_2) \derives b.}\\
\end{array}$\\
$\begin{array}{ll}
\bf (5) & \bf projection \ rules :\\
&\begin{array}{l@{\qquad}l}
\irule{a \derives a'(X).} & \irule{\tneg a \derives \tneg a'(X).}\\
\irule{b \derives b'(X).} & \irule{\tneg b \derives \tneg b'(X).}\\
\irule{c \derives c'(X).} & \irule{d \derives d'(X).}\\
\irule{e \derives e'(X).} & \\
 \end{array}
\end{array}$\\
$\begin{array}{ll}
\bf (6) & \bf constraints:\\
&\irule{\derives  a, \tneg a.}\\
&\irule{\derives  b, \tneg b.}
\end{array}$\\
}
$\punto$
\end{example}

Given a model $M$ for $DLP(\gp)$, $\pi(M)$ is the set of literals
obtained from $M$ by eliminating all the literals with a ``primed''
predicate symbol, i.e. a predicate symbol in the set $\{ prec', ovr'
\} \cup \{ \phi'  \ |  \ \exists \ an \ adorned \ literal$
$\phi(\bar X) \ appearing \ in \ \gp \}$. $\pi(M)$ is the set of literals
without all atoms which were introduced by the translation algorithm.

The DLP version of a $\DLPI$-program $\gp$ can be used in place of
$\gp$ in order to evaluate answer sets of $\gp$.  The result
supporting the above statement is the following:
\begin{theorem}\label{equivalence}
Let $\gp$ be a $\DLPI$-program. Then,
for each answer set $M$ for $\gp$ there exists a consistent answer set
$M'$ for $DLP(\gp)$ such that $\pi(M') = M$.
Moreover, for each consistent answer set $M'$ for $DLP(\gp)$ there exists
an answer set $M$ for $\gp$ such that $\pi(M') = M$.
\end{theorem}

\begin{proof}
First we show that given an answer set $M$ for $\gp$ there exists
a consistent answer set $M'$ for $DLP(\gp)$ such that $\pi(M') = M$.
We proceed by constructing the model $M'$.
Let $K_1 = \{ prec'(o,o_1) \mid o < o_1 \}$.
Let $K_2$
be the set of ground literals
$ovr'(r,o, \bar X)$
such that there exists a (defeasible) rule $r \in ground(\gp)$
with $obj\_of(r) = o$ such that $r$ is overridden in $M$
and $\bar X$ is the tuple of arguments appearing in
the head of $r$.
Let $K_3$ be the set of ground literals
$ovr'(L,o)$ such that there exist two rules
$r,r' \in ground(\gp)$ such that
$L \in Head(r)$, $r$ is defeasible and $r'$ overrides
$r$ in $L$.
Let denote by $\cal K$ the collection of sets
of ground literals such that each element
$K \in \cal K$ satisfies the following properties:
\begin{enumerate}
\item
for each literal $\phi(\bar X) \in M$ there is a
literal $\phi'(o,\bar X)$ in $K$, for some object identifier $o$,
\item
for each $r \in G_M(\gp)$ such that the body of $r$ is true in
$M$, for at least one literal $\phi(\bar X)$ of the head of $r$ a
corresponding literal $\phi'(obj\_of(r),\bar X)$ occurs in $K$,
\item
$K \subseteq \{ \phi'(o,\bar X) \ | \ \phi(\bar X) \
\mbox{is an adorned predicate appearing in}\  \gp \wedge o \in {\cal O} \}$,
\item
$K$ is a consistent set of literals.
\end{enumerate}
First observe that
the family $\cal K$ is not empty
(i.e., there is at least a set of consistent sets of literals
satisfying items (1), (2) and (3) above).
This immediately follows from the fact
that $M$ is an answer set of the program $\gp$.

Let $M_{K} = K_1 \cup K_2 \cup K_3 \cup K \cup M$,
for a generic $K \in \cal K$. It is easy to
show that $G_{M_{K}}(DLP(\gp))$ is independent on which $K \in
\cal K$ is chosen. Indeed, no literals from $K$ appear in the NAF
part of the rules in $ground(DLP(\gp))$.

Now we examine which rules the program
$pos(G_{M_{K}}(DLP(\gp)))$ contains
(for any set $K \in \cal K$).

Both the rules with head predicate $prec'$ and $ovr'$ and the rules of
the form $\phi(\bar X) \derives \phi'(X, \bar X)$ (added by
statement 21 of Figure \ref{alg1}) appear unchanged in
$pos(G_{M_K}(DLP(\gp)))$. Indeed, these rules do not contain a NAF
part (recall that the GL transformation can modify only rules in which a
NAF part occurs). Each constraint of $DLP(\gp)$ (added by statement 26
of the algorithm), that is a rule of the form $b \derives \phi(\bar
X), \tneg . \phi(\bar X), \dneg \ b$ (where $b$ is a literal not
occurring in $M_K$) is translated into the rule $b \derives \phi(\bar
X), \tneg . \phi(\bar X)$.

The other rules in $pos(G_{M_K}(DLP(\gp)))$ originate from
rules of $ground(DLP(\gp))$ obtained by rewriting rules of
$ground(\gp)$ (see statements 11 and 15 of the algorithm).

Thus, consider a rule $r$ of $ground(\gp)$.

If $r$ is defeasible and overridden in $M$ then the
corresponding rule in $ground(DLP(\gp))$
(generated by statement 11 of the algorithm) contains a NAF part not
satisfied in $M_K$, by construction of $K_2$ and $K_3$. Hence, such
a rule appears neither in $pos(G_M(\gp))$ nor in $pos(G_{M_K}(DLP(\gp)))$.

The other case we have to consider
is that the rule $r$ is either a strict rule
or a defeasible rule not overridden in $M$.

First suppose that $r$ is a strict rule or is a defeasible rule
not threatened in $M$ (recall that a rule not threatened in $M$ is
certainly not overridden in $M$).
In this case, the corresponding rule, say $r'$
in $ground(DLP(\gp))$ (generated by
statement 15 of the algorithm)
has the same body of $r$
and the head modified by renaming predicates
(from $\phi$ to $\phi'$) and by adding the object $o$
(from which the rule $r$ comes) as first
argument in each head literal.
Since the body of $r'$ does not contain literals
from $K_1, K_2, K_3$ and $K$,
and further $M \subseteq M_K$ (for each $K \in \cal K$),
$r'$ is eliminated by the GL transformation w.r.t. $M_K$ if and only
if $r$ is eliminated by the GL transformation w.r.t. $M$.
Moreover, in case $r'$ is not eliminated by the GL transformation
w.r.t $M_K$,
since the body of $r'$ does not contain literals
from $K_1, K_2, K_3$ and $K$,
the GL transformation w.r.t. $M_K$ modifies
the body of $r'$ in the same way
the GL transformation w.r.t. $M$ modifies the body
of $r$.
Thus, each rule $r$ in
$pos(G_M(\gp))$ has a corresponding rule in $pos(G_{M_K}(DLP(\gp)))$ with
the same body and a rewritten head.

Now suppose that $r$ is a threatened defeasible rule
that is not overridden in $M$.
In this case, the corresponding rule, say $r'$
in $ground(DLP(\gp))$ (generated by
statement 11 of the algorithm)
has the head modified by renaming predicates
(from $\phi$ to $\phi'$) and by adding the object $o$ as first
argument in each head literal
and a body obtained by adding
to the body of $r$ a literal of
the form $\dneg  \ ovr'(r,o, \bar X)$, where
$o$ is the object from which $r$ comes, and $\bar X$
represents the tuple of terms appearing in the
head literals of $r$.
Since the rule $r$ is
not overridden in $M$, the literal $ovr'(r,o, \bar X)$
cannot belong to $K_2$ and hence cannot
belong to $M_K$.
Thus, the GL transformation w.r.t $M_K$ eliminates
the NAF part of the rule $r'$.
As a consequence, also in this case, each rule in
$pos(G_M(\gp))$ has a corresponding rule in $pos(G_{M_K}(DLP(\gp)))$ with

Now we prove that, for any $K \in \cal K$,
$M_K$ is a model for
$pos(G_{M_K}(DLP(\gp)))$.
Indeed, rules with head predicate $prec'$
are clearly satisfied.
Further, rules with head predicate $ovr'$
are satisfied by construction of set $K_2$,
$K_3$ and $K$.
Moreover, rules of the form
$\phi(\bar X) \derives \phi'(X, \bar X)$ are
satisfied since $M \subseteq M_K$ and
by construction of $K$.
Rules of $pos(G_{M_K}(DLP(\gp)))$
of the form $b \derives \phi(\bar X), \dneg \ \phi(\bar X)$,
originated by the translation of the constraints,
are satisfied since $M_K$ is a consistent set of literals.

Consider now the remaining rules
(those corresponding to rules of $pos(G_M(\gp))$).
Let $r$ be a rule of $pos(G_{M_K}(DLP(\gp)))$
and $r'$ the rule of
$pos(G_M(\gp))$ corresponding to $r$.
As shown earlier, the two rules have
the same body.
Thus, if the body of $r$ is true
w.r.t. $M_K$, the body of $r'$ is
true w.r.t. $M$, since no literal
of $M_K \setminus M$ can appear in the body
of the rule $r'$ (and hence of the rule $r$).
As a consequence, by property (2) of
the collection $\cal K$ to which the
set $K$ belongs, the head of the rule $r$
is true in $M_K$.

Thus, $M_K$ is a model
for $pos(G_{M_K}(DLP(\gp)))$.

Now we prove the following claim:

{\bf Claim 1.} Let $\bar M$ be a model for $pos(G_{M_K}(DLP(\gp)))$.
Then, $\pi(\bar M)$ is a model for
$pos(G_{M}(\gp))$.

\begin{proof}
By contradiction suppose that $\pi(\bar M)$ is not a model
for  $pos(G_{M}(\gp))$. Thus, there exists a rule $r \in  pos(G_{M}(\gp))$
with body true in $\pi(\bar M)$ and head false in $\pi(\bar M)$.
Since, as shown earlier, the rule $r$ has a corresponding
rule $r'$ in  $pos(G_{M_K}(DLP(\gp)))$ with the same body
of $r$ and the head obtained by replacing each literal
$\phi(\bar X)$ of the head of $r$ by the corresponding literal
$\phi'(o,\bar X)$, where $o$ is the object from which $r$ comes.
Since  $\pi(\bar M)$ is a model for $pos(G_{M_K}(DLP(\gp)))$,
at least one of the "$\phi'$ literal"  of the head of $r'$ must be
true in $\pi(\bar M)$. Then, due to the presence
of the rules of the form $\phi(\bar X) \derives \phi'(X, \bar X)$ in
$pos(G_{M_K}(DLP(\gp)))$,  $\pi(\bar M)$ must contain also the "$\phi$
corresponding literal" belonging to the head of $r$ (contradiction)
\end{proof}

Moreover we  prove that each model
for $pos(G_{M_K}(DLP(\gp)))$
\begin{description}
\item(1)
contains $M$, and
\item(2)
belongs to $\cal K$.
\end{description}
To prove item (1), suppose by contradiction $\bar M$ is
a model for $pos(G_{M_K}(DLP(\gp)))$ such that
$M \cap \bar M \neq M$.
Thus, $\pi(\bar M) \subset M$.
On the other hand, by Claim 1, $\pi(\bar M)$
is a model for $pos(G_M(\gp))$.
But since $M$ is an answer set for $\gp$
and then a minimal model for
$pos(G_M(\gp))$, a contradiction arises.

Now we have to prove the item (2) above. 
First observe that
the properties 3. and 4. of the family $\cal K$ are trivially verified
by the models of $pos(G_{M_K}(DLP(\gp)))$.
Thus, by contradiction suppose there exists a model
$\bar M$ of $pos(G_{M_K}(DLP(\gp)))$
such that it does not satisfy
one of the properties 1. or 2. characterizing the family $\cal K$. 

First suppose that property 1. is not satisfied by $\bar M$, that is,
there is a literal $\phi (\bar X)$ in $M$ such that no
corresponding literal $\phi'(o, \bar X)$ occurs in $\bar M$, for
some object identifier $o$. Let $\bar M'$ be the set obtained by
$\bar M$ by eliminating all such literals $\phi(\bar X)$. It is
easy to see that $\bar M'$ is still a model for
$pos(G_{M_K}(DLP(\gp)))$. Indeed, rules of the form $\phi(\bar X)
\derives \phi'(X, \bar X)$ are satisfied since literals $\phi(\bar
X)$ dropped from $\bar M$ do not have corresponding "$\phi'$
literals" by hypothesis. Further, no other rule in
$pos(G_{M_K}(DLP(\gp)))$ contains a literal from $M$ in the head.
On the other hand, by Claim 1, $\pi(M')$ is a model
for $pos(G_M(\gp))$. But this is a contradiction,
since $\pi(M') \subset M$ and $M$ is an answer set for $\gp$.

Suppose now that property 2. is not satisfied by $\bar M$,
that is, there is a rule $r \in pos(G_M(\gp))$ such
that the body of $r$ is true in $M$ and
no "$\phi'$ literals"  corresponding
to literals of the head occur in $\bar M$.
Since, $M \subseteq \bar M$ (see item (1) above), the corresponding
rule of $pos(G_{M_K}(DLP(\gp)))$ is not satisfied.
But this implies that $\bar M$ can not be a model
for $pos(G_{M_K}(DLP(\gp)))$ (contradiction).

A consequence of the fact that every model of
$pos(G_{M_K}(DLP(\gp)))$ contains $M$ is that
every model of $pos(G_{M_K}(DLP(\gp)))$ must
contain the sets $K_1$, $K_2$ and $K_3$.
because of the rules added by statements
1,5 and 12 of the algorithm.

Thus, the model
$M'=M_{K'}$ for $K' \in \cal K$ such
that no set
$\bar K \in \cal K$ exists
such that $\bar K \subset K'$
is a minimal model for
$pos(G_{M_K}(DLP(\gp))$, that is,
a consistent answer set for $DLP(\gp)$.
Hence,
the first part of the proof is concluded, since
$\pi(M') = M$.

Now, we prove that
given a consistent answer set $M'$ for $DLP(\gp)$,
$M = \pi(M')$ is an answer set for $\gp$.

First we prove that a literal $\phi(\bar X)$ belongs
to $M$ if and only if
there exits a literal $\phi'(o, \bar X)$ in
$M'$, for some object identifier $o$.

Indeed, $\phi(\bar X) \in M$ implies that
$\phi(\bar X) \in M'$. But since $M'$ is a minimal model
for $pos(G_{M'}(DLP(\gp))$, there must exits a rule in
$pos(G_{M'}(DLP(\gp))$ with head
containing the literal $\phi(\bar X)$ and body
true w.r.t. $M'$ (otherwise the literal
$\phi(\bar X)$ could be dropped from
$M'$ without invalidate any rule of
$pos(G_{M'}(DLP(\gp))$ and thus $M'$ would not
be minimal).
Conversely, if $\phi'(o, \bar X) \in M'$, 
for some object identifier $o$, the literal
$\phi(\bar X)$ belongs to $M'$, since $M'$
is a model for $pos(G_{M'}(DLP(\gp))$
and the rule $\phi(\bar X) \derives \phi'(o, \bar X)$ belongs
to $pos(G_{M'}(DLP(\gp))$.
Thus,  $\phi(\bar X) \in M$.

Moreover, we prove that every rule of
$pos(G_{M'}(DLP(\gp))$ has a corresponding
rule in $pos(G_M(\gp))$ with same body
and a head obtained by replacing the $\phi'$
literals with the $\phi$ corresponding ones
and by eliminating the object argument
from these literals.
Indeed, from the above result, the GL transformation
deletes a rule $r$ from $ground(DLP(\gp))$ if either
the corresponding rule belonging to $ground(\gp)$
is overridden in $M$ (due the the literal
$\dneg  \ ovr'(r,o,\bar X)$ occurring in the body of $r$)
or some negated (by negation $\dneg$) literal is
false in $M'$. But this literal is false in
$M'$ if and only if it is false in $M$.
On the other hand, in case the rule is not
deleted by the GL transformation, its body
is rewritten in the same way of the corresponding
rule appearing in $pos(G_M(\gp))$.

As a consequence, $M$ is a model for $pos(G_{M}(\gp))$.
Indeed, if the body of a rule $r$ of
$pos(G_{M}(\gp))$ is true w.r.t. $M$, the corresponding
rule $r'$ of  $pos(G_{M'}(DLP(\gp))$ has the body true
w.r.t. $M'$. Hence, at least one of the head literals
of $r'$ must be true in $M'$.
Let $\phi'(o, \bar X)$ such a literal.
As shown earlier, this implies that $\phi(\bar X)$
belongs to $M$.
But $\phi(\bar X)$ appears in the head of $r$
and hence $r$ is satisfied in $M$.

Now we prove that $M$ is minimal.
By contradiction, suppose that $\bar M \subset M$ is
a model for $pos(G_{M}(\gp))$.
Consider the literals belonging to the set
$M  \setminus \bar M$. Because
of the correspondence between the rules
of $pos(G_{M}(\gp))$ and the rules
of $pos(G_{M'}(DLP(\gp))$, the set of literals
obtained from $M'$ by eliminating
all the literals $\phi(\bar X)$ belonging
to the set $M \setminus \bar M$ as well as
the corresponding $\phi'$ literals
is still a model for $pos(G_{M'}(DLP(\gp))$.
But this is a contradiction,
since $M'$ is a consistent answer set for
$DLP(\gp)$.

Since $M$ is a model for  $pos(G_{M}(\gp))$
and is minimal, $M = \pi(M')$ is an answer set
for the program $\gp$. Hence the proof is concluded.
\end{proof}

\begin{example}\label{ex-equivalence}
{
Consider the program $\gp$ of Example \ref{ex-1}.
It is easy to see that
$DLP(\gp)$ (see Example \ref{ex-translation}) admits
one consistent answer set
$M = \{prec'(o_2,o_1), a'(o_1), e'(o_1),$
$b'(o_2),$ $c'(o_2),$ $ovr'(\tneg b,o_1),$ $a,$ $e,$ $b,$ $c \}$.
Thus, $\pi(M) = \{a,b,c,e\}$.
On the other hand,
$\pi(M)$ is the only answer
set for $\gp$,
as shown in Example \ref{ex-4}.
}
$\punto$
\end{example}

\subsection{System Architecture}

\begin{figure}
\epsfig{file=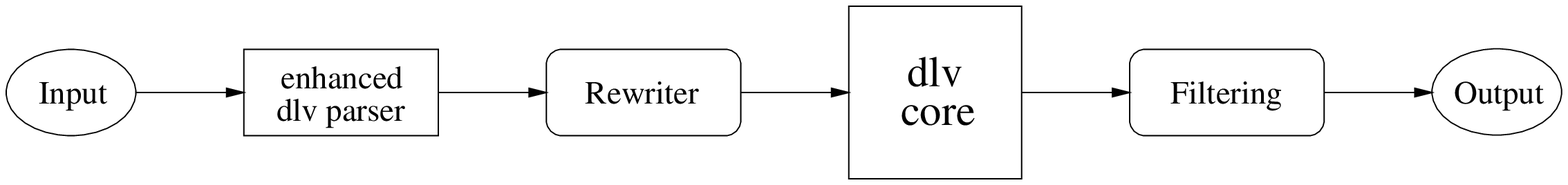,width=\textwidth}
\caption{Flow diagram of the system.}
\label{architecture}
\end{figure}

We have used the \dlv system \cite{eite-etal-98a} to implement a system
for $\DLPI$. The concept is that of a front-end to plain DLP, which
has been used before for implementing various ways of reasoning modes
and languages on top of the \dlv system. The front-end implements the
translation described in Section~\ref{sec:implementation}. A schematic
visualization of its architecture is shown in Figure
\ref{architecture}.

First of all, we have extended the \dlv parser to incorporate the
$\DLPI$ syntax. In this way, all the advanced features of \dlv
(e.g.\ bounded integer arithmetics, comparison built-ins, etc.) are also
available with $\DLPI$.
The {\em Rewriter} module implements the translation depicted
in Figure \ref{alg1}.
Once the rewritten version $\pi(\gp )$ of $\gp$ is generated,
its answer sets
are then computed using the \dlv core. Before the output is
shown to the user, $\pi$ is applied to each answer set in order to strip the
internal predicates from the output.

On the webpage \cite{dlvi-web} the system is described in detail. It
has been fully incorporated into the \dlv system. To use it, just
supply an input file using the syntax described in
Section~\ref{syntax} --- \dlv automatically invokes the $\DLPI$
frontend in this case.

Note that it is currently required to specify the objects in the order
of the inheritance hierarchy: Specifying that some object inherits
from another object which has not been defined before will result in
an error. Since cyclic dependencies are not allowed in our language,
this requirement is not a restriction.

\section{Conclusion}\label{sec:conclusion}

We have presented a new language, named $\DLPI$,
resulting from the extension of (function-free)
disjunctive logic programming with inheritance.
$\DLPI$ comes with a declarative model-theoretic semantics,
where possible conflicts are solved in favor of more specific
rules (according to the inheritance hierarchy).
$\DLPI$ is a consistent generalization of the Answer Set Semantics
of disjunctive logic programs, it respects some fundamental
inheritance principles, and it seems to be suitable
also to give a semantics to updates in logic programs. 

While inheritance enhances the knowledge representation
modeling features of disjunctive logic programming,
the addition of inheritance does not increase
its computational complexity.
Thus, inheritance ``comes for free'': the user can
take profit of its knowledge modeling ability,
without paying any extra cost in terms of computational load.
It was therefore possible to implement a $\DLPI$ system
on top of the disjunctive logic programming system \dlv.
The system is freely available on the web \cite{dlvi-web} and ready-to-use
for experimenting with the use of inheritance in KR applications.

\section*{Acknowledgements}

The idea of representing inertia in planning problems in a higher
object is due to Axel Polleres. The introduction of the notion of
strict rules was suggested by Michael Gelfond. This work was
partially supported by FWF (Austrian Science Funds) under the
projects \mbox{P11580-MAT} and \mbox{Z29-INF}.


\bibliography{bibtex}
\end{document}